\documentclass[aps,prl,twocolumn,groupedaddress,noshowpacs,checklength,amsmath]{revtex4-1}

\usepackage{graphicx}
\usepackage{amssymb}
\usepackage{amsmath}
\usepackage{cancel}

\def\ihat{{\bf \hat x}}
\def\jhat{{\bf \hat y}}
\def\khat{{\bf \hat z}}

\begin{document}

\title{Comment on ``Trouble with the Lorentz Law of Force: Incompatibility with Special Relativity and Momentum Conservation"}

\author{David J.~Griffiths}\email[Electronic address: ]{griffith@reed.edu}
\affiliation{Department of Physics, Reed College, Portland, Oregon  97202}
\author{V.~Hnizdo}
\affiliation{National Institute for Occupational Safety and Health, Morgantown, West Virginia 26505}

\begin{abstract}
A recent article claims the Lorentz force law is incompatible with special relativity.  The claim is false, and the ``paradox" on which it is based was resolved many years ago.
\end{abstract}

\maketitle

In a recent article in this journal \cite{Man}, Mansuripur claims to present ``incontrovertible theoretical evidence of the incompatibility of the Lorentz [force] law with the fundamental tenets of special relativity," and concludes that ``the Lorentz law must be abandoned."  His argument is based on the following thought experiment:

Consider an ideal magnetic dipole ${\bf m} = m_0\,\ihat$, a distance $d$ from a point charge $q$, both of them at rest  (Figure 1).  The torque on {\bf m} is obviously zero.  Now examine the same system from the perspective of an observer moving to the left at constant speed $v$.  In this reference frame \cite{lab} the (moving) point charge generates electric and magnetic fields
\begin{eqnarray}
{\bf E}(x,y,z,t) &= &\frac{q}{4\pi\epsilon_0}\frac{\gamma}{R^3}\left(x\,\ihat + y\,\jhat + (z-vt)\,\khat\right),\\
{\bf B}(x,y,z,t) &= &\frac{q}{4\pi\epsilon_0}\frac{v\gamma}{c^2R^3}\left(-y\,\ihat + x\,\jhat\right),
\end{eqnarray}
($\gamma \equiv 1/\sqrt{1-(v/c)^2}$, $R\equiv \sqrt{x^2+y^2+\gamma^2(z-vt)^2}\,$), and the (moving) magnetic dipole acquires an electric dipole moment \cite{VH}
\begin{equation}
{\bf p} = \frac{1}{c^2}({\bf v}\times {\bf m})=\frac{1}{c^2}vm_0\,\jhat.
\end{equation}
The net torque on the electric/magnetic dipole is
\begin{equation}
{\bf N} = ({\bf m}\times {\bf B}) + ({\bf p}\times {\bf E})  = \frac{qm_0}{4\pi\epsilon_0}\frac{v}{c^2d^2}\,\ihat
\end{equation}
(by Lorentz transformation, $d =z'= \gamma(z-vt)$; the magnetic contribution is zero, because {\bf B} vanishes on the $z$ axis).   Apparently the torque on the dipole is zero in one inertial frame, but {\it non}-zero in another!  Mansuripur concludes that the Lorentz force law (on which Eq.~4 is predicated) is inconsistent with special relativity.

\vskip0in
\begin{figure}[t]
\hskip-.2in\scalebox{.7}[.7]{\includegraphics{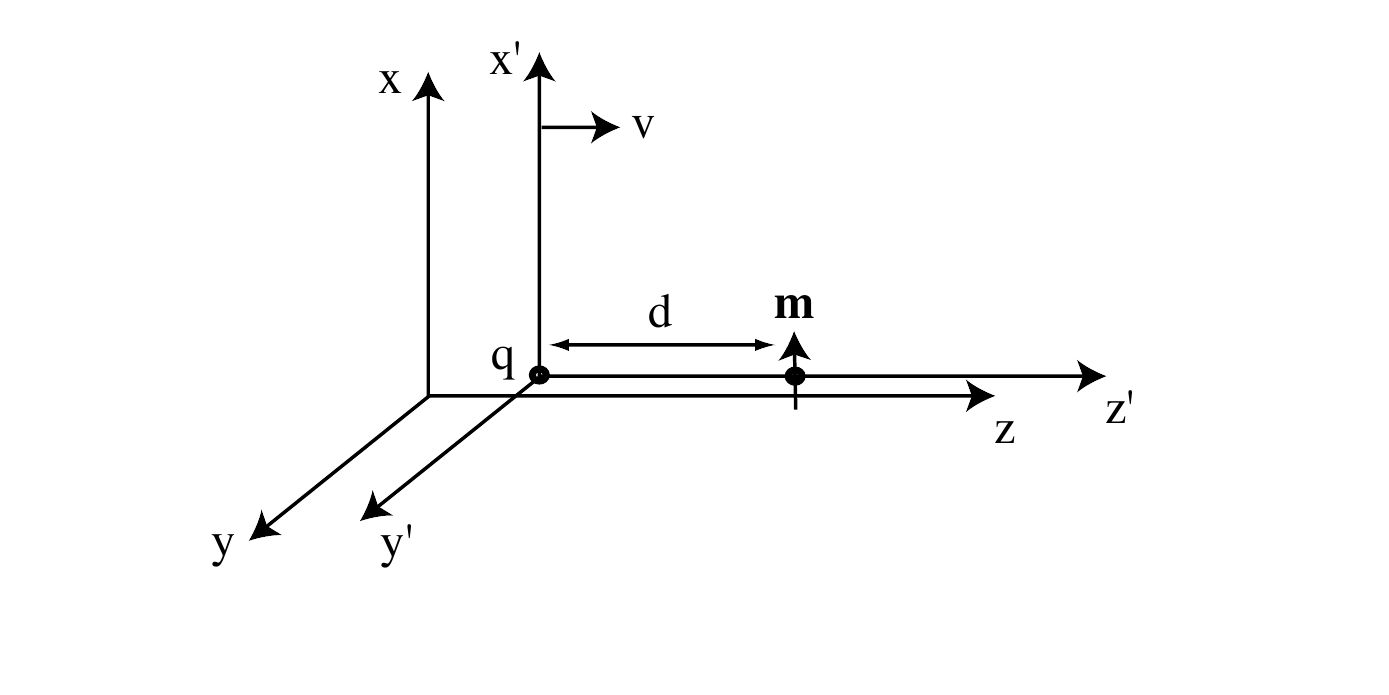}}
\vskip-.3in
\caption{Electric charge ($q$) and magnetic dipole ({\bf m} in proper (primed) and lab (unprimed) frames.}
\end{figure}

This ``paradox" was resolved many years ago by Victor Namias \cite{Namias}.  He noted that the standard torque formulas (${\bf p}\times {\bf E}$ and ${\bf m}\times {\bf B}$) apply to dipoles {\it at rest}.  Suppose we model the magnetic dipole as separated north and south poles.   The ``Lorentz force law" for a magnetic monopole $q^*$ reads \cite{DJG1}
\begin{equation}
{\bf F} = q^*\left({\bf B} - \frac{1}{c^2}{\bf v}\times {\bf E}\right),
\end{equation}
so the torque \cite{origin} (in the lab frame) on a moving dipole ${\bf m}=q^*{\bf s}$ is
\begin{eqnarray}
{\bf N} &=& ({\bf r}_+\times {\bf F}_+)+({\bf r}_-\times {\bf F}_-)\nonumber\\
&=&({\bf r}_+-{\bf r}_-)\times \left[q^*\left({\bf B}-\frac{1}{c^2}{\bf v}\times {\bf E}\right)\right]\nonumber\\
&=&q^*{\bf s}\times\left({\bf B}-\frac{1}{c^2}{\bf v}\times {\bf E}\right)\nonumber\\
&=& ({\bf m}\times {\bf B})-\frac{1}{c^2}{\bf m}\times({\bf v}\times {\bf E}).
\end{eqnarray}
Using the vector identity ${\bf m}\times({\bf v}\times {\bf E})= {\bf v}\times({\bf m}\times {\bf E})+({\bf m}\times{\bf v})\times {\bf E}$,
\begin{equation}
{\bf N} = ({\bf m}\times {\bf B})-\frac{1}{c^2}({\bf m}\times{\bf v})\times {\bf E} - \frac{1}{c^2}{\bf v}\times({\bf m}\times{\bf E}).
\end{equation}
In view of Eq.~3,
\begin{equation}
{\bf N} = ({\bf m}\times{\bf B})+({\bf p}\times{\bf E})-\frac{1}{c^2}{\bf v}\times({\bf m}\times {\bf E}).
\end{equation}
According to Namias, then, Eq.~4 is simply in error---there is a third term, which, it is easy to check, exactly cancels the offending torque.  The {\it net} torque, correctly calculated, is zero in both reference frames \cite{orientation}.

Namias believed that his formula (Eq.~7) applies just as well to an ``Amp\`ere dipole" (an electric current loop) as it does to a ``Gilbert dipole" (separated magnetic monopoles) \cite{Namias_quote}.   He was {\it almost} right.  As we have known since the 1960's, an Amp\`ere dipole in an electric field carries ``hidden" momentum
\cite{hid mom},
\begin{equation}
{\bf p}_h = \frac{1}{c^2}({\bf m}\times {\bf E}).
\end{equation}
(This is in the rest frame of the dipole, but since in our case it is perpendicular to {\bf v}, it is correct as well in the lab frame.)    The hidden momentum is constant, because the charge is a fixed distance from the dipole; there is therefore no associated force.  On the other hand, the hidden {\it angular} momentum,
\begin{equation}
{\bf L}_h = {\bf r}\times{\bf p}_h,
\end{equation}
is {\it not} constant (in the lab frame), because {\bf r} is changing.  In fact,
\begin{equation}
\frac{d{\bf L}_h}{dt} = {\bf v} \times {\bf p}_h = \frac{1}{c^2}{\bf v}\times({\bf m}\times {\bf E}).
\end{equation}

Torque is the rate of change of angular momentum:
\begin{equation}
{\bf N} = \frac{d{\bf L}}{dt} = \frac{d{\bf L}_o}{dt}+\frac{d{\bf L}_h}{dt},
\end{equation}
where ${\bf L}_o$ is the ``overt" angular momentum (associated with actual rotation of the object), and ${\bf L}_h$ is the ``hidden" angular momentum (so called because it is {\it not} associated with any overt rotation of the object).
If we pull the second term over to the left \cite{logic}:
\begin{equation}
{\bf N}_{\rm eff}\equiv{\bf N} - \frac{d{\bf L}_h}{dt} = \frac{d{\bf L}_o}{dt}.
\end{equation}
It is this ``effective" torque that Namias obtains in Eq.~7.  Physically $d{\bf L}_h/dt$ is not a torque, but (the rate of change of) a piece of the angular momentum \cite{fictitious}.  The torque itself is given (in the Amp\`ere model) by Eq.~4; on the other hand, it is the ``overt" torque that must vanish to resolve the paradox, since the dipole is not rotating \cite{Bedford}.   In the Gilbert model the torque is given by Eq.~7, the total is zero, and the angular momentum is constant; in the Amp\`ere model the torque is given by Eq.~4, and it accounts for the change in (hidden) angular momentum \cite{lab_torque}.

It is of interest to see exactly how this plays out in Mansuripur's formulation of the problem.  He treats the dipole as magnetized medium, and calculates the torque directly from the Lorentz force law, without invoking ${\bf p}\times {\bf E}$ or ${\bf m}\times {\bf B}$.  In the proper frame, he takes
\begin{equation}
{\bf M}'(x',y',z',t')= m_0\delta(x')\delta(y')\delta(z'-d)\,\ihat.
\end{equation}
Now, {\bf M} and {\bf P} constitute an antisymmetric second-rank tensor; transforming to the lab frame we find
\begin{eqnarray}
{\bf M}(x,y,z,t)&= &m_0\delta(x)\delta(y)\delta\left(z-vt-(d/\gamma)\right)\,\ihat,\\
{\bf P}(x,y,z,t)&=& \frac{m_0v}{c^2}\delta(x)\delta(y)\delta\left(z-vt-(d/\gamma)\right)\,\jhat.
\end{eqnarray}

According to the Lorentz law, the force density is
\begin{equation}
{\bf f} = \rho {\bf E} + {\bf J}\times{\bf B},
\end{equation}
where $\rho = -\nabla\cdot {\bf P}$ is the bound charge density and ${\bf J} = \partial {\bf P}/\partial t + \nabla \times{\bf M}$ is the sum of the polarization current and the bound current density.  Using Eqs.~1,2,15, and 16, we obtain
\begin{eqnarray}
{\bf f} &=& -(\nabla\cdot {\bf P}){\bf E} +(\nabla \times {\bf M})\times {\bf B} + \frac{\partial {\bf P}}{\partial t}\times{\bf B}\nonumber\\
&=&-\frac{qm_0v}{4\pi\epsilon_0 c^2}\frac{d}{R^3}\delta(x)\delta'(y)\delta(z-vt-d/\gamma)\,\khat.
\end{eqnarray}
The net force on the dipole is
\begin{equation}
{\bf F} = \int{\bf f}\,dx\,dy\,dz = \frac{qm_0vd}{4\pi\epsilon_0 c^2}\frac{d}{dy}\left[\frac{1}{(y^2+d^2)^{3/2}}\right]\Bigg|_{y=0}\,\khat =  {\bf 0}.
\end{equation}
Meanwhile, the torque density is
\begin{equation}
{\bf n} = {\bf r}\times {\bf f} = -\frac{qm_0vd}{4\pi\epsilon_0 c^2}\frac{y}{R^2}\delta(x)\delta'(y)\delta(z-vt-d/\gamma)\,\ihat,
\end{equation}
and the net torque on the dipole is
\begin{eqnarray}
{\bf N} &=& \int {\bf n}\,dx\,dy\,dz\nonumber\\
&=&-\frac{qm_0vd}{4\pi\epsilon_0 c^2}\left\{-\frac{d}{dy}\left[\frac{y}{(y^2+d^2)^{3/2}}\right]\right\}\Bigg|_{y=0}\,\ihat\nonumber\\
&=&\frac{qm_0v}{4\pi\epsilon_0 c^2d^2}\,\ihat,
\end{eqnarray}
confirming Eq.~4.  This is precisely the torque required to account for the increase in hidden angular momentum.

What if we run Mansuripur's calculation for a dipole made out of magnetic monopoles?  In that case there will be no hidden momentum to save the day.  Well, the bound charge, bound current, and magnetization current are \cite{minus}
\begin{equation}
\rho^*_b = -\nabla\cdot {\bf M},\quad {\bf J}^*_b = -c^2\nabla\times {\bf P},\quad {\bf J}^*_p=\frac{\partial{\bf M}}{\partial t},
\end{equation}
so the force density on the magnetic dipole (again invoking Eqs.~1, 2, 15, and 16) is
\begin{eqnarray}
{\bf f}&=& \rho^*{\bf B} -\frac{1}{c^2}{\bf J}^*\times{\bf E} \nonumber\\
&=& -(\nabla \cdot {\bf M}){\bf B}+(\nabla \times{\bf P})\times{\bf E} - \frac{1}{c^2}\left(\frac{\partial {\bf M}}{\partial t}\right)\times {\bf E}\nonumber\\
&=& {\bf 0}.
\end{eqnarray}
The total force is again zero, but this time so too is the torque density (${\bf n} = {\bf r}\times {\bf f}$), and hence the total torque.

{\it Conclusion.}  The resolution of Mansuripur's ``paradox" depends on the model for the magnetic dipole:
\begin{itemize}
\item If it is a Gilbert dipole (made from magnetic monopoles), the third term in Namias' formula (Eq.~7) supplies the missing torque.  In Mansuripur's formulation, in terms of a polarizable medium, it comes from a correct accounting of the bound charge/current (Eq.~22).  The net torque is zero in the lab frame, just as it is in the proper frame.
\item If it is an Amp\`ere dipole (an electric current loop), the third term in Namias' equation is absent, and the torque on the dipole is {\it not} zero.  It is, however, just right to account for the increasing hidden angular momentum in the dipole.
\end{itemize}
In either model the Lorentz force law is entirely consistent with special relativity \cite{postings}.

We thank Kirk McDonald for useful correspondence.   VH coauthored this comment in his private capacity; no official support or endorsement by the Centers for Disease Control and Prevention is intended or should be inferred.


\begin{thebibliography}{99}
\bibitem{lab} We'll call it the ``lab" frame, and call the rest frame of the charge and dipole the ``proper" frame (following Mansuripur, we use primed coordinates for the latter).
\bibitem{fictitious} As Namias suggests, this approach is analogous to the introduction of ``fictitious" forces (centrifugal, Coriolis, \ldots) in classical mechanics, when using an accelerating reference frame.
\end{thebibliography}
\end{document}